# Coronagraph-integrated wavefront sensing with a sparse aperture mask

Hari Subedi
Neil T. Zimmerman
N. Jeremy Kasdin
Kathleen Cavanagh
A J Eldorado Riggs

# Coronagraph-integrated wavefront sensing with a sparse aperture mask

**Hari Subedi,* Neil T. Zimmerman, N. Jeremy Kasdin, Kathleen Cavanagh, and A J Eldorado Riggs**
Princeton University, Department of Mechanical and Aerospace Engineering, Engineering Quadrangle, Olden Street, Princeton, New Jersey 08544, United States

**Abstract.** Stellar coronagraph performance is highly sensitive to optical aberrations. In order to effectively suppress starlight for exoplanet imaging applications, low-order wavefront aberrations entering a coronagraph, such as tip-tilt, defocus, and coma, must be determined and compensated. Previous authors have established the utility of pupil-plane masks (both nonredundant/sparse-aperture and generally asymmetric aperture masks) for wavefront sensing (WFS). Here, we show how a sparse aperture mask (SAM) can be integrated with a coronagraph to measure low-order differential phase aberrations. Starlight rejected by the coronagraph's focal plane stop is collimated to a relay pupil, where the mask forms an interference fringe pattern on a subsequent detector. Our numerical Fourier propagation models show that the information encoded in the fringe intensity distortions is sufficient to accurately discriminate and estimate Zernike phase modes extending from tip-tilt up to radial degree $n = 5$, with amplitude up to $\lambda/20$ RMS. The SAM sensor can be integrated with both Lyot and shaped pupil coronagraphs at no detriment to the science beam quality. We characterize the reconstruction accuracy and the performance under low flux/short exposure time conditions, and place it in context of other coronagraph WFS schemes.  

## 1 Introduction

High-contrast, direct-imaging techniques enable astronomers to detect exoplanets and characterize their atmospheres and orbits. Recent progress in this area from ground-based observations has included the measurement of molecular absorption features in the near-infrared spectra of the planets in the HR 8799 system, and the young Jovian analog GJ 504 b.[1–4] The essential challenge of exoplanet imaging is to suppress starlight at small angular separations, thereby achieving a region of high contrast in the image where the light of a planet is detected.

Coronagraphs manipulate the diffraction of starlight to enable the extreme dynamic range necessary to directly image exoplanets. The shaped pupil coronagraph (SPC) is one of several families of solutions.[5] The baseline architecture of the wide-field infrared survey telescope–astrophysics focused telescope assets[6] (WFIRST-AFTA) mission includes an SPC, and those designs are currently undergoing tests at the Jet Propulsion Laboratory's high contrast imaging testbed facility.[7–9] The SPC applies an optimized binary apodization at the re-imaged telescope pupil to create a region of high contrast in the image plane.

The sensitivity of coronagraph performance to low-order wavefront error has been described before for apodized pupil Lyot coronagraphs (APLCs) and phase-induced amplitude apodization coronagraphs.[10,11] The SPC also requires the phase in the plane of the apodizer to be tightly controlled. We demonstrate an example in Fig. 1, for an SPC with a contrast goal below $10^{-9}$. First, the point spread function (PSF) is shown with an ideal, flat wavefront, and then with low-order phase aberrations added. In this case, two Zernike modes were added: defocus and coma modes, each with root-mean square (RMS) amplitude $\lambda/30$. The plot in the right panel shows how the resulting distortion worsens the intensity pattern by over a factor of 100 at separation $5\lambda/D$.

Even in the absence of an atmosphere, the wavefront error contributions from imperfect optical surfaces, mechanical stresses, and thermal distortions of the primary mirror and supporting structures can easily exceed the aberration requirements of a coronagraph. Left uncorrected, they cause unwanted starlight to leak into the dark regions of the image intended for planet detection. Line-of-sight pointing jitter is another source of concern for exoplanet imaging instruments both on the ground and in space. In addition to starlight leaking into the region of scientific interest, if jitter is severe enough then electric field estimation will be corrupted, limiting the efficacy of high-order wavefront control. Therefore, in addition to low-order modes such as astigmatism and coma originating within the instrument, it is crucial to rapidly monitor tip-tilt errors entering the coronagraph. In some cases, structural vibrations can also demand the compensation of focusing errors on short time scales.

It is possible to correct the wavefront entering the coronagraph based on science image plane measurements and surface modulations applied to the deformable mirror(s), in effect solving for a "dark hole" actuator solution[12] via speckle nulling[13,14] or electric field conjugation.[15–17] Also, the method of phase diversity provides a way to estimate static phase errors from the science image plane.[18,19] However, all these image plane methods are blind to short time-scale variations,[20] and lack the core of the star's PSF whence low-order modes can be

*Address all correspondence to: Hari Subedi, E-mail: hsubedi@princeton.edu

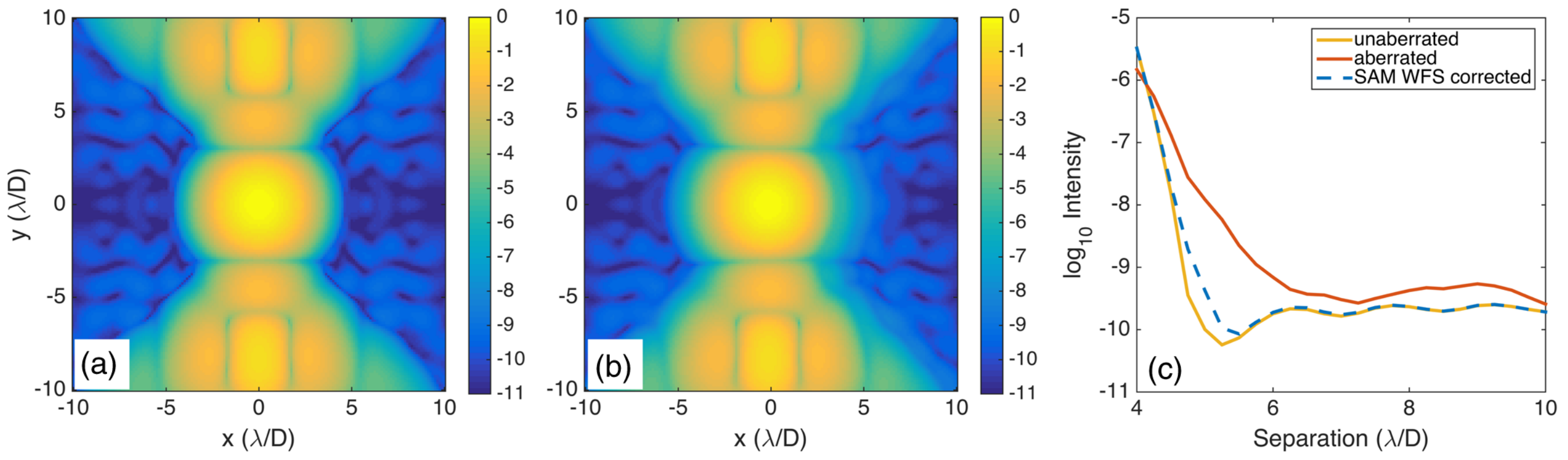

**Fig. 1** (a) Nominal (flat wavefront) point spread function (PSF), plotted on a log scale; (b) point spread function with defocus and coma modes added to the wavefront, each with strength $\lambda/30$ RMS waves; (c) azimuthally averaged contrast curve for cases of flat wavefront, defocus and coma modes, and after removing the aberration estimated by the SAM WFS model.

most easily inferred. They also require lengthy calibration procedures exclusive of science exposure time. Therefore, a number of sensing schemes have been proposed that rely on auxiliary hardware outside the coronagraph to augment the atmospheric correction provided by an adaptive optics (AO) system (in the case of ground-based observatories), as well as to reduce the demand placed on the much slower image plane calibration loop. These methods have various limitations in aberration sensitivity and a wide range in implementation complexity; we briefly survey their properties.

Early efforts combining Lyot coronagraphy with AO used discarded starlight from the focal plane mask (implemented as a through hole in a flat mirror) to sense tip-tilt errors.[21,22] The coronagraphic low-order wavefront sensor (CLOWFS) presented by Guyon et al.[11] builds on this approach to sense additional aberration modes. In the CLOWFS scheme, the coronagraph focal plane mask is an oblique transmissive plane with an opaque occulting spot. The outer ring of the occulting spot is reflective, directing light from an annulus of the star's PSF (with inner and outer radii ∼0.7 and $1.8\lambda/D$) to a lens outside of the science path. Guyon et al. place a charge-coupled device (CCD) detector in the converging beam after this lens, but offset slightly from its focus. Tip, tilt, defocus, and two orthogonal astigmatism modes are fitted to the critically sampled intensity distribution in the defocused image.[11,23] The displacement of the sensing plane from true focus is necessary to retrieve the defocus component of the wavefront when the magnitude of that error is in the small phase regime (≪1 radian). In fact, it can be shown with additional analysis that estimating any phase mode with even symmetry requires the CLOWFS sensor to be defocused.[24]

Other solutions for coronagraph wavefront sensing (WFS) rely on external interferometric calibration.[25–28] One approach in this category that avoids some of the complexities typically associated with constructing an interferometer is the phase-contrast technique. Zernike first developed the phase contrast sensor in 1933 to convert optical path differences in translucent microscope specimens to intensity signals.[29] It was proposed for astronomical applications by Dicke,[30] and since then the concept has been adapted for coronagraph WFS[31] and evaluated on several instrument testbeds.[9,32,33] A phase-contrast wavefront sensor [also known as a Zernike wavefront sensor (ZWFS)] also uses the light extracted from the occulted region of the focal plane as in CLOWFS. The core of the beam is then focused on a $\pi/2$ phase-shifting spot of diameter $\sim 1-2\ \lambda/D$. This phase-shifting spot can also be built into the focal plane mask of the coronagraph to eliminate unnecessary optics. For small phase errors, the intensity of the interference pattern in the subsequent pupil is a linear function of the phase distribution in the original pupil.[34] The baseline architecture of the WFIRST-AFTA mission includes a ZWFS for sensing the low-order wavefront aberrations.[6]

Outside of coronagraphy, aperture masking with nonredundant or sparse aperture masks has carved compelling niches in diffraction-limited, high dynamic range astronomical observations.[35–37] In this approach, a binary mask is inserted in a relay pupil plane, consisting of a set of subaperture holes arranged so that the baselines between each pair of holes are unique—hence the term nonredundant.[38–40] The beam is then brought to a focus on a detector forming an interference pattern. Fourier analysis of the resulting image plane interference pattern can retrieve closure phase observables that are invariant to time-varying low-order aberrations, and otherwise inaccessible with conventional filled aperture imaging.[41] It has been recognized that the wavefront information encoded in aperture mask data also provides powerful diagnostics of systematic AO residuals,[42,43] and provides a way to robustly cophase the mirror segments of a space telescope.[44,45] Martinache recently extended aperture mask WFS to operate with generally asymmetric aperture geometries, rather than purely nonredundant masks.[46] This technique, so far, is used on the Subaru coronagraphic extreme AO (SCExAO) instrument[47] and on the PALM-3000 extreme AO system on the Palomar 200-inch telescope.[48] Due to the need to have a full PSF for this technique, neither of the configurations had a coronagraph.

In our present work, we show that SAM WFS is an attractive solution to the dynamic low-order aberration correction requirements of a coronagraph. It is well-suited for space coronagraph applications, where small differential aberrations from some starting point need to be canceled during the operation of a slow, high-order wavefront control loop. Like the CLOWFS, and the Zernike phase mask sensor for the WFIRST/AFTA, our SAM WFS method relies on starlight discarded by the focal plane mask, and is therefore minimally invasive to the science beam. Similiarly to the nonredundant methods, we have a SAM at the relay pupil. Instead of doing a Fourier analysis of the image plane interference pattern, we use a linear equation to relate the fringe pattern in the sensor to the aberrations in the pupil plane. Using the monochromatic propagation model, linearity response of differential low-order phase aberrations up to scale $\lambda/20$ RMS is studied.

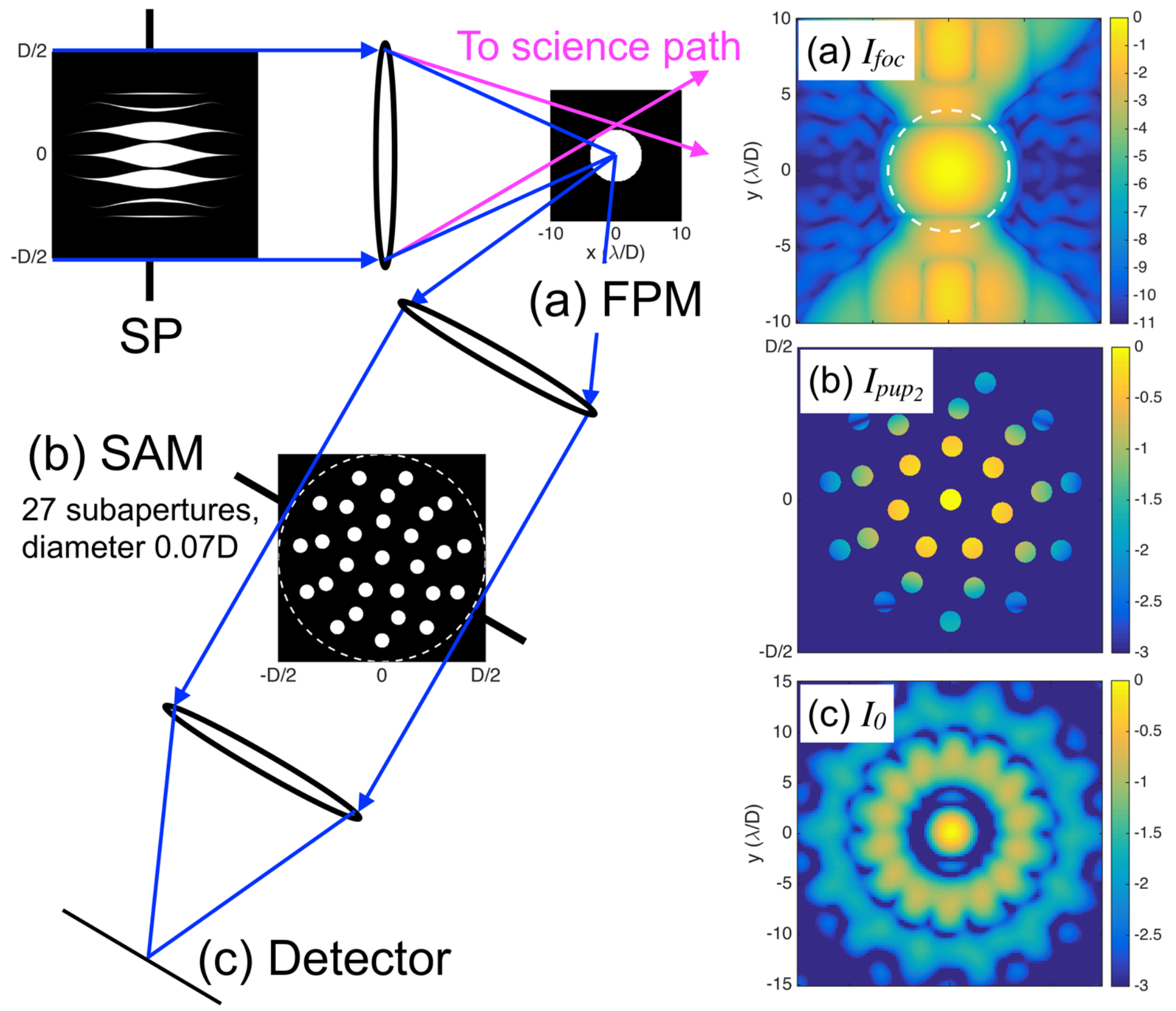

**Fig. 2** Diagram of the sparse aperture mask (SAM) wavefront sensor (WFS), integrated with a shaped pupil coronagraph (SPC). On the right side, we plot the intensity at three critical planes on a logarithmic flux scale: the coronagraph focal plane, the sparse aperture mask, and the sensor.

We begin in Sec. 2 with an overview of SAM WFS design and a mathematical description of the linear estimation system. In Sec. 3, we describe our Fourier propagation model and several performance trials. Finally, we discuss the performance in the context of other coronagraph WFS schemes and plans for future work.

## 2 Optical Configuration

A diagram of the SAM WFS is given in Fig. 2. Incoming starlight, apodized by the shaped pupil, is brought to a focus at a focal plane mask. This oblique mask reflects light from the core of the PSF toward the SAM WFS and transmits the remaining light to the science camera optics. The collimating optic following the reflective focal plane mask forms a re-imaged pupil at the SAM. The light diffracted by the pattern of holes in the SAM is brought to a focus on a dedicated detector. The right-hand side of Fig. 2 depicts the intensity pattern at the three critical planes: the first focal plane of the coronagraph, the SAM plane, and the sensor focus.

The nominal physical parameters of our propagation model are summarized in Table 1. We chose the pupil diameter and focal length to match the planned experimental setup at the Princeton High Contrast Imaging Laboratory.

### 2.1 *Focal Plane Mask*

The reflective focal plane mask directs discarded starlight light toward the WFS. The radius of this region can be as small as the first zero of the PSF core, or as wide as the inner working angle of the coronagraph permits. For the model presented in this paper, we fix the radius at $4\lambda/D$, to match the inner working angle of one of our ripple SPC designs.

Unlike conventional coronagraph apodizations, a shaped pupil can produce a PSF with significant energy outside the main lobe in certain directions. This is because the shaped pupil optimization procedure can be tailored to create dark search zones restricted to a finite region of the image plane, in exchange for deeper contrast and small inner working angle.[7,49] This is the case for the ripple SPC, as evident by the vertically extended sidelobe wedges of the nominal star PSF shown in Fig. 1. Therefore, in addition to the simple reflecting spot, we tested a variation on the focal plane mask to utilize the starlight outside the central lobe of the PSF. The only modification, as illustrated in Fig. 3, is an extension of the reflective region to an 95-deg wide wedge above and below the center, creating a "bowtie"-like shape out to a separation of $10\lambda/D$ (Fig. 3).

**Table 1** Nominal physical parameters of the wavefront sensor (WFS) optical propagation model.

| | |
|---|---|
| Pupil diameter | 10 mm |
| Pupil sampling | 512 points/diameter |
| Focal length of imaging optics | 200 mm |
| Wavelength | 550 nm |
| Spatial sampling at sensor image plane | 4 pixels per $\lambda/D$ |
| Sensor image width | 80 pixels (20 $\lambda/D$) |

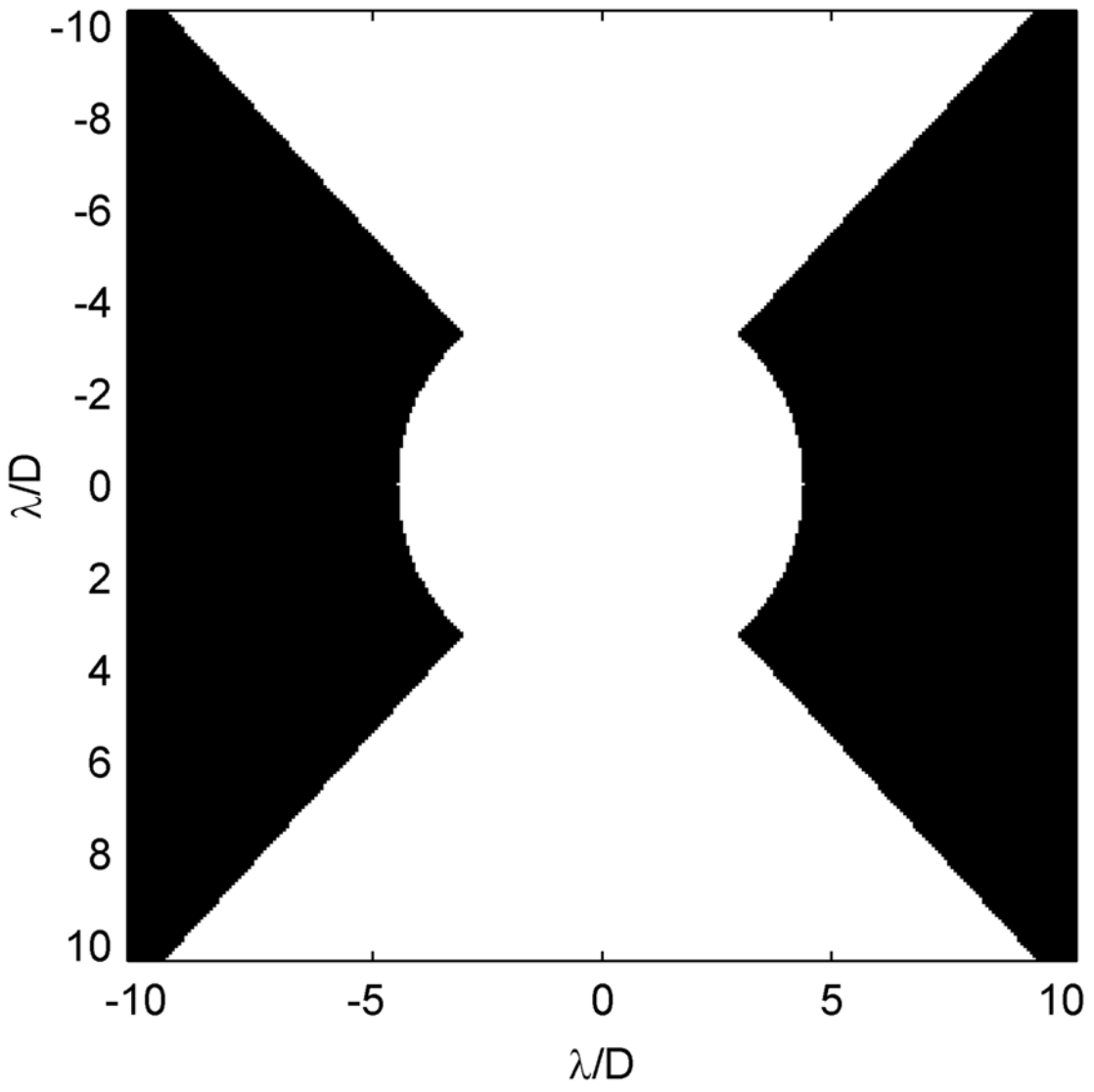

**Fig. 3** Diagram of the alternative focal mask design for the SAM WFS, a reflective "bowtie" taking advantage of the shape of the PSF created by the ripple-shaped pupil.

## 2.2 *Sparse Aperture Mask*

The light reflected from the focal plane mask is collimated to a relay pupil where the WFS aperture mask is located. The starlight re-imaged after propagating through the mask produces an interference pattern characteristic of the baselines between each pair of holes. The difference between the aberrated and nominal intensity patterns—shown in Fig. 4 for three wavefront examples—provides the estimation signal. For this illustration, we normalize the nominal intensity pattern (labeled $I_0$ in Fig. 2) to unity to show the relative scale of the differential signal. Table 2 provides the peak differential intensity for the first 14 Zernike aberrations after piston.

The aperture mask $\mathcal{M}$, over a two-dimensional (2-D) Cartesian vector $\mathbf{u}$ in the telescope pupil plane (with the origin at the center of the pupil), is defined as in Zimmerman,[50] according to

$$\mathcal{M}(\mathbf{u}) = \Pi(\mathbf{u}/a) * \sum_{i=1}^{N_h} \delta(\mathbf{u} - \mathbf{h_i}), \tag{1}$$

where $N_h$ is the number of holes, $\mathbf{h_i}$ is the vector position of the center of each subaperture, $\Pi$ is the subaperture hole function, and $\delta$ is the Dirac delta function. For circular subapertures of radius $a$, $\Pi$ is defined via

$$\Pi(\mathbf{u}/a) = \begin{cases} 1, & \text{if } |\mathbf{u}|/a \leq 1 \\ 0, & \text{otherwise.} \end{cases} \tag{2}$$

**Table 2** Peak differential intensity $(I_{ab} - I_0)/I_0$ of SAMWFS for a fixed $\lambda/30$ RMS wavefront error.

| Noll index (after piston Z1) | Peak differential intensity |
|---|---|
| 2 | $5.42 \times 10^{-2}$ |
| 3 | $5.34 \times 10^{-2}$ |
| 4 | $7.58 \times 10^{-3}$ |
| 5 | $8.37 \times 10^{-3}$ |
| 6 | $7.31 \times 10^{-3}$ |
| 7 | $1.03 \times 10^{-1}$ |
| 8 | $1.02 \times 10^{-1}$ |
| 9 | $4.74 \times 10^{-3}$ |
| 10 | $3.96 \times 10^{-3}$ |
| 11 | $1.92 \times 10^{-2}$ |
| 12 | $1.88 \times 10^{-2}$ |
| 13 | $2.57 \times 10^{-2}$ |
| 14 | $1.72 \times 10^{-3}$ |
| 15 | $1.78 \times 10^{-3}$ |

The design parameters of the aperture mask are the number, sizes, and locations of the subaperture holes. The need to sense even-symmetric Zernike modes such as defocus requires the pattern to be asymmetric with respect to the center of the pupil. Beyond this constraint, there is a large parameter space to explore. For the present demonstration, we tailored the SAM design for the ripple shaped pupil shown in Fig. 2, aiming to recover Zernike phase modes of radial degree up to $n = 4$ (14 modes in total, excluding piston). All the quantitative results presented in this paper are based on the SAM design shown in Fig. 2, which uses 27 subaperture holes, each with a diameter 7% of the pupil. This specific mask is intended only to serve as a proof of concept. The numerical optimization of the SAM geometry will be a subject of a separate study. We add further remarks on SAM design considerations in the discussion in Sec. 4.

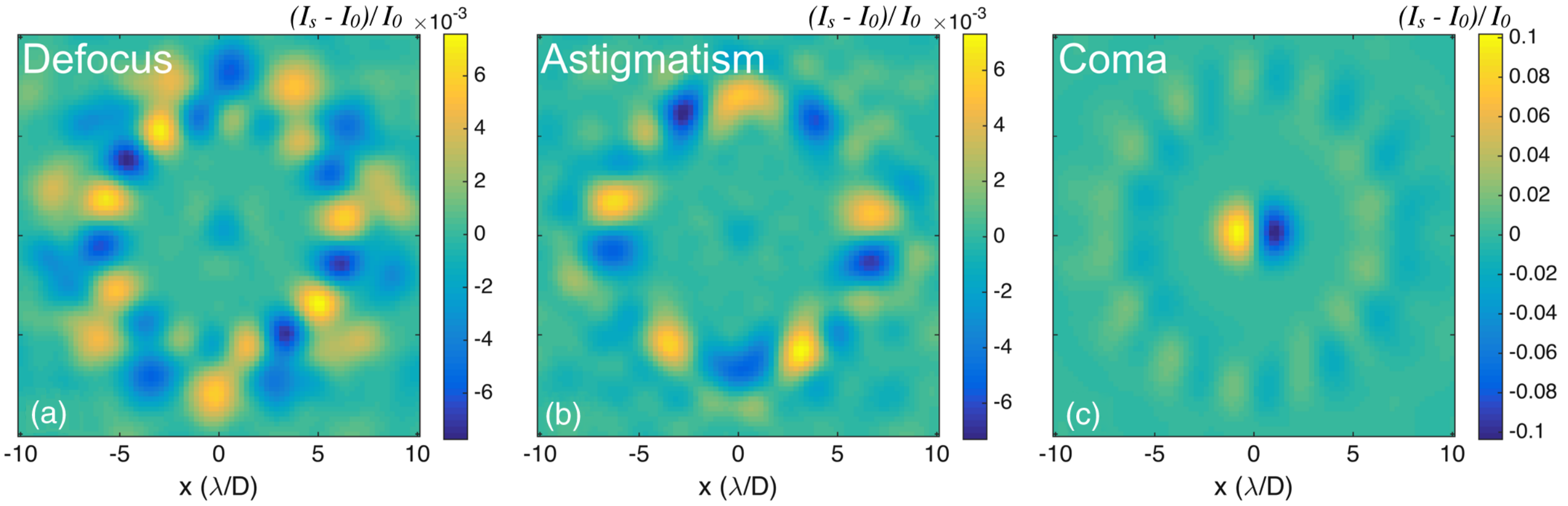

**Fig. 4** The differential intensity patterns for three aberration cases: (a) defocus ($\lambda/30$ waves RMS), (b) astigmatism ($\lambda/30$ waves RMS), and (c) coma ($\lambda/30$ waves RMS).

### 2.3 *Sparse Aperture Mask Sensing Equation*

In this section, we develop an equation to relate the fringe pattern in the sensor to the aberrations in the pupil plane. In the small aberration regime, the field at the first focal plane $E_{\text{foc}}$ is the Fourier transform of Eq. (4) given by

$$\begin{aligned} E_{\text{foc}}(\nu) &= \mathcal{F}[A(r,\theta)e^{i\phi(r,\theta)}] \\ &\simeq \mathcal{F}[A(r)] + i\sum_{nm} a_{nm}\mathcal{F}[A(r)Z_n^m(r,\theta)]. \end{aligned} \tag{3}$$

Here, $Z_n^m(r,\theta)$ is the Zernike polynomial of radial order $n$ and azimuthal order $m$, defined on a circle of unity radius and normalized to unity variance. The orders $m$ and $n$ are consolidated to Noll index $k$ and the Zernike polynomials are now written as $Z_k$.

The binary-valued focal plane mask $\mu_{\text{SAM}}(\mathbf{x})$ reflects some of the light to the WFS, where $\mathbf{x}$ is a 2-D Cartesian vector in the focal plane. The arguments $r$, $\theta$, $\mathbf{u}$, and $\mathbf{x}$ are dropped hereafter to simplify the reading. The field at the relay pupil plane following the reflection, $E_{\text{pup}_2}$, is given by the Fourier transform of the product of the mask and the field at the focal plane such that

$$\begin{aligned} E_{\text{pup}_2} &= \mathcal{F}(\mu_{\text{SAM}}E_{\text{foc}}) \\ &= \mathcal{F}[\mu_{\text{SAM}}\mathcal{F}(A)] + i\sum_k a_k\mathcal{F}[\mu_{\text{SAM}}\mathcal{F}(AZ_k)]. \end{aligned} \tag{4}$$

The field at the sensor plane, obtained after the light passing through the SAM is focused by an optic, is then given by the product of the SAM transmission function and the field at the relay pupil plane. The field at the sensor plane is

$$\begin{aligned} E_s &= \mathcal{F}(\mathcal{M}E_{\text{pup}_2}) \\ &= \mathcal{F}\{\mathcal{M}\mathcal{F}[\mu_{\text{SAM}}\mathcal{F}(A)]\} + i\sum_k a_k\mathcal{F}\{\mathcal{M}\mathcal{F}[\mu_{\text{SAM}}\mathcal{F}(AZ_k)]\}. \end{aligned} \tag{5}$$

This field can be represented as $E_s = E_0 + E_{ab}$ where the nominal field $E_0$ is given by

$$E_0 = \mathcal{F}\{\mathcal{M}\mathcal{F}[\mu_{\text{SAM}}\mathcal{F}(A)]\}, \tag{6}$$

and the effect of the aberrations on the field, $E_{ab}$, is given by

$$E_{ab} = i\sum_k a_k\mathcal{F}\{\mathcal{M}\mathcal{F}[\mu_{\text{SAM}}\mathcal{F}(AZ_k)]\}. \tag{7}$$

The intensity at the sensor plane is given by

$$\begin{aligned} I_s &= E_s\bar{E}_s \\ &= (E_0 + E_{ab})(\bar{E}_0 + \bar{E}_{ab}) \\ &= E_0\bar{E}_0 + 2\,\text{Re}(E_0\bar{E}_{ab}) + E_{ab}\bar{E}_{ab} \\ &\simeq I_0 + 2\text{Re}(E_0\bar{E}_{ab}), \end{aligned} \tag{8}$$

where $I_0$ is the nominal intensity at the sensor plane and $E_{ab}\bar{E}_{ab}$ is neglected because $a_k^2 \ll 1$. Therefore, the sensing equation is linear and given by

$$(I_s - I_0)/2 = \text{Re}\{E_0\bar{E}_{ab}\}. \tag{9}$$

Substituting Eq. (7) into Eq. (9), and rearranging the 2-D electric field distribution of each mode into a column vector,

$$\begin{aligned} (I_s - I_0)/2 &= \text{Re}\{E_0\bar{E}_{ab}\} \\ &= \sum_k \text{Re}\{E_0\overline{ia_k\mathcal{F}\{\mathcal{M}\mathcal{F}[\mu_{\text{SAM}}\mathcal{F}(AZ_k)]\}}\} \\ &= [\text{Re}\{E_0\overline{i\mathcal{F}\{\mathcal{M}\mathcal{F}[\mu_{\text{SAM}}\mathcal{F}(AZ_2)]\}}\}\ldots\ldots] \begin{bmatrix} a_2 \\ a_3 \\ \cdot \\ \cdot \\ a_k \end{bmatrix}. \end{aligned} \tag{10}$$

Equation (10) can be written as

$$(I_s - I_0)/2 = Hx, \tag{11}$$

where the response or the modal matrix

$$H = [\,\text{Re}\{E_0\overline{i\mathcal{F}\{\mathcal{M}\mathcal{F}[\mu_{\text{SAM}}\mathcal{F}(AZ_2)]\}}\} \quad \ldots \quad \ldots], \tag{12}$$

and $x = [a_j]$, $j = 1\ldots k$. The modal matrix $H$ and the nominal image $I_0$ are calculated analytically based on the optical configuration. The modal matrix can also be obtained by propagating different Zernike modes through the system and observing the resulting images at the detector. The solution to Eq. (11) is then obtained using a least-squares fit.

As CLOWFS is one of the experimentally verified, mature, low-order WFSs, and its optical configuration is similar to what we have designed for SAM WFS; therefore, it is used as a reference point for comparison purposes.

### 2.4 *Coronagraphic Low-Order Wavefront Sensor Sensing Equation*

The CLOWFS devised by Guyon et al. is similar to the SAM WFS, in that it also relies on re-imaged starlight discarded by the focal plane stop.[11] However, there are three essential differences between the CLOWFS and the SAM WFS. First, for CLOWFS, the focal plane region reflected to the sensor is annular so as to exclude the central core of the PSF. Second, rather than collimating the extracted light to form a re-imaged pupil, the light is simply directed to a single converging lens or mirror. Finally, the detector in the CLOWFS sensing plane must be offset from true focus of the last converging optic, to enable estimation of aberration modes with even symmetry.[11,24]

In the view of using CLOWFS as a benchmark for our concept, here we develop an analogous sensing equation for that system, which we later employ in our numerical model. Optically, the field at the CLOWFS sensor is equivalent to the field in the plane at a small distance after the reflective focal plane mask $\mu_{\mathrm{CS}}$. This defocused field is computed by the Fresnel integral, which we abbreviate as the operator $\mathcal{F}r\{\,\}$. We again symbolize the reflective annulus by the binary variable $\mu_{\mathrm{CS}}$. Then, analogous to Eq. (5), for the CLOWFS sensor electric field we have

$$\begin{aligned} E_s &= \mathcal{F}r\{\mu_{\mathrm{CS}}\mathcal{F}[A(r,\theta)e^{i\phi(r,\theta)}]\} \\ &= \mathcal{F}r[\mu_{\mathrm{CS}}\mathcal{F}(A)] + i\sum_k a_k \mathcal{F}r[\mu_{\mathrm{CS}}\mathcal{F}(AZ_k)] \\ &= E_0 + E_{ab}. \end{aligned} \tag{13}$$

Using the same approximation as in Eq. (8), dropping the $E_{ab}\bar{E}_{ab}$ term in the intensity expression, we again find a system of linear equations relating the aberration coefficients to the differential intensity pattern:

$$\begin{aligned} (I_s - I_0)/2 &= \mathrm{Re}\{E_0\bar{E}_{ab}\} \\ &= \sum_k \mathrm{Re}\{E_0\overline{ia_k\mathcal{F}r[\mu_{\mathrm{CS}}\mathcal{F}(AZ_k)]}\} \\ &= [\mathrm{Re}\{E_0\overline{i\mathcal{F}r[\mu_{\mathrm{CS}}\mathcal{F}(AZ_2)]}\}\ldots\ldots]\begin{bmatrix} a_2 \\ a_3 \\ \cdot \\ \cdot \\ a_k \end{bmatrix}. \end{aligned} \tag{14}$$

To obtain the aberration coefficients for CLOWFS same approach from Sec. 2.3 was used.

## 3 Simulation and Analysis

Our numerical Fourier propagation model of the SAM WFS uses a 512-point diameter representation of the ripple shaped pupil shown in Fig. 2. We sample the focal planes (both coronagraph and sensor) with eight points across each $\lambda/D$ resolution element. As mentioned in Sec. 2, we test two focal plane mask shapes for the SAM WFS: a disk of radius $4\lambda/D$, and a disk with "bowtie" wedges (Fig. 3). The array representing the CCD in the sensor plane bins the intensity points $2 \times 2$, resulting in a final image resolution of four pixels per $\lambda/D$. Bit quantization, photon counting noise, and three analog-to-digital units (ADU) read noise are then applied to give a simulated intensity image.

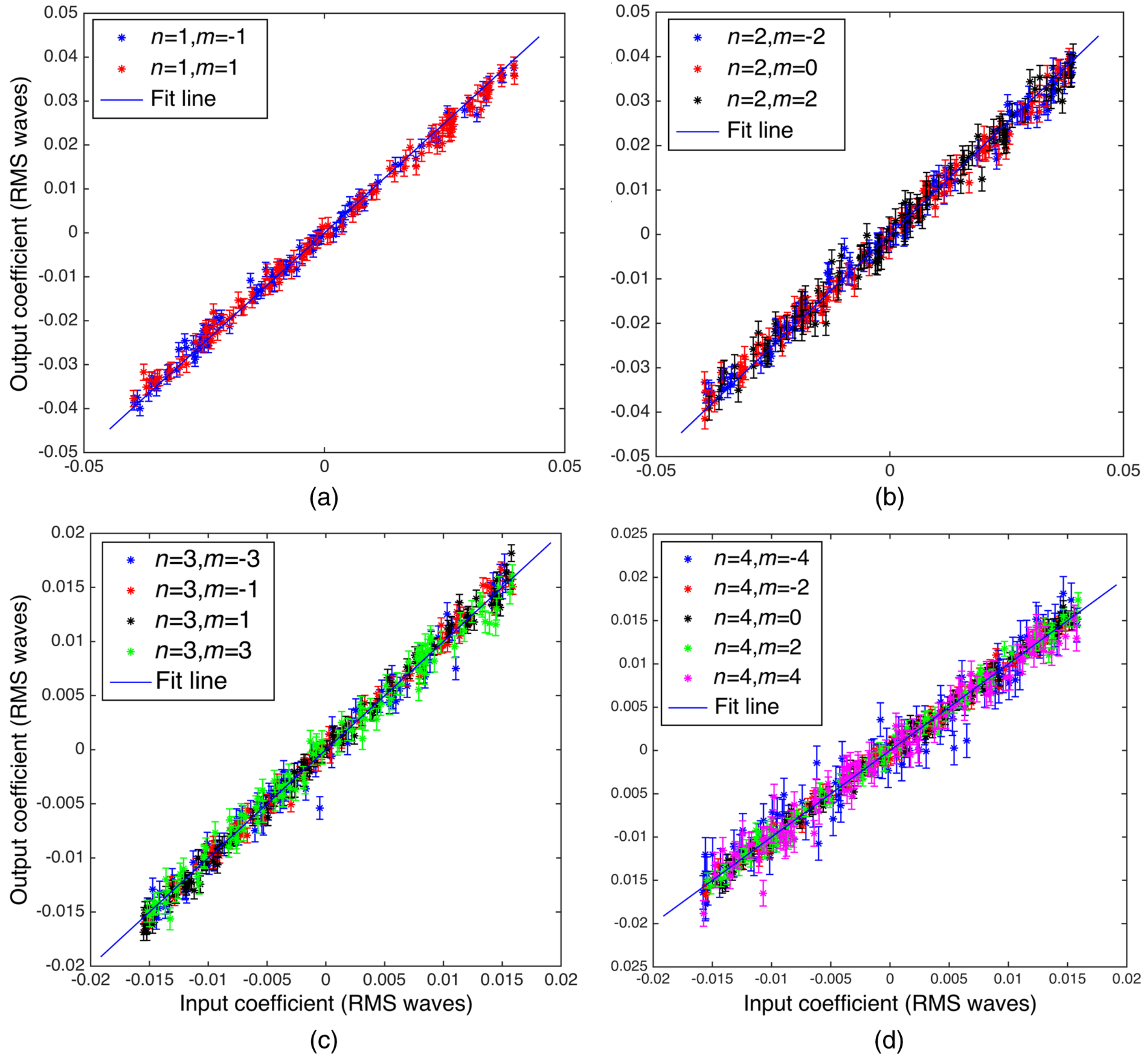

**Fig. 5** Inferred versus input coefficients, in units of waves RMS for (a) Zernike radial degree $n = 1$, (b) radial degree $n = 2$, (c) radial degree $n = 3$ and (d) radial degree $n = 4$.

The CLOWFS propagation model is similar, but requires a near-field Fresnel integral to compute the defocused sensor field after the coronagraph focal plane. The Fresnel propagation requires us to define several physical dimensions: pupil diameter, focal length, propagation distance (defocus), and wavelength. For these we used the same parameters as for our planned testbed, listed in Table 1. The outer radius of the reflective focal annulus was fixed at $4\lambda/D$, identical to the SAM WFS. We then tuned both the defocus and the inner radius of the annulus by trial and error to give the best performance in terms of $R^2$, a metric described in Sec. 3.1. For our coronagraph and range of mode estimation, we found that the CLOWFS worked best without any inner radius, instead leaving the reflective region as a simple disk. We found the best performance at a defocus at 3% of the 200 mm focal length, or 6 mm. For our 550 nm wavelength, and 10 mm diameter pupil, this corresponds to a defocus aberration of 1.88 $\mu$m peak to valley, equivalent to 3.4 waves. We note this is very close to the 3.3 waves used by Guyon et al. in their published design.[11] For both SAM and CLOWFS simulations, the same stellar magnitude, spectral bandwidth, photon counting noise, read noise, and integration time were used.

### 3.1 *Fit Analysis*

To evaluate the SAM WFS concept, we constructed 100 aberrated wavefront realizations from random combinations of Zernike polynomials. This was repeated for two ranges of mode input and estimation: Zernike radial degree $n = 1 - 4$, and $n = 1 - 5$. The polynomial coefficients were drawn from uniform distributions, with larger bounds for the lower order terms than the higher-order terms: 0.04 waves RMS for $n = 1$ and 2, 0.016 waves RMS for larger radial degree. These coefficients are all small enough to satisfy the linearity assumption. Since lower-order modes such as tip-tilt and defocus are strongest and quickly varying, we chose a higher value for these modes ($n = 1 - 2$). Nevertheless, the propagation model used the full expression for the field at the each plane, rather than the first-order expansion used for the estimation equation. In these trials, the pixel with the maximum intensity was assumed to have 40,000 counts (near full-well on a 16-bit CCD) while other pixels receive photons based on their relative intensity.

We plot the estimated coefficients versus the input coefficients of these trials in Fig. 5, for the case of the SAM WFS with a circular focal plane mask, estimating modes up to radial degree 4. The tight clustering of all these points to the $y = x$ line is a good indication of the accuracy of the sensor, and indicates that there is no significant cross-talk or degeneracy between modes in the system.

In Table 3, we collect the mean estimation errors across the ensemble of wavefront realizations, for each WFS configuration, and for both ranges of mode estimations. The phase is compared only within the open area of the shaped pupil. For a typical wavefront in the $n = 1 - 4$ trial ensemble, the subtracted SAM WFS estimate reduces the RMS wavefront error by a factor of 30, and the peak to valley by a factor of 10. The estimate with the bowtie focal plane mask (FPM) is only marginally better than the circular case. For the CLOWFS, the RMS residuals are typically a factor of 2 to 3 worse than the SAM. For the $n = 1 - 5$ trial, SAM estimates reduce the typical RMS error by a factor of 20, and the improvement offered by the bowtie FPM over the circular FPM remains slight.

We also analyzed the random wavefront estimates with the coefficient of determination method ($R^2$ fit analysis). The fit assessment value was obtained by

$$R^2 = 1 - \frac{\text{RSS}}{\text{TSS}}. \tag{15}$$

Residual sum of squares (RSS) is the sum of the squared errors of the inferred mode coefficients versus "truth" over all wavefront realizations, measuring the discrepancy between the data and the estimation model. Total sum of squares (TSS) is the sum of the squared differences of the inferred coefficients from their respective means over all realizations. An aberration mode is considered to be estimated well if $R^2 > 0.5$. Both WFSs maintain $R^2$ values well above this cutoff for all phase modes (Fig. 6). However, the accuracy advantage of SAM in this regime of full-well signal is apparent by the fact that lowest $R^2$ value for SAM is 0.95 with an average $R^2$ value of 0.97, versus 0.79 for CLOWFS with an average of 0.9.

**Table 3** Mean residual errors of the estimates of 100 random wavefronts, as measured within the open area of the shaped pupil. For each WFS configuration and radial degree estimation range, we give the mean RMS residual, and the mean peak-to-valley residual. The top row gives the mean wavefront errors for the uncorrected input realizations. All values are in units of wavelength.

| | $n = 1 - 4$ | | $n = 1 - 5$ | |
|---|---|---|---|---|
| | RMS in | $P - V$ in | RMS in | $P - V$ in |
| | 0.015 | 0.097 | 0.018 | 0.11 |
| | RMS residual | $P - V$ residual | RMS residual | $P - V$ residual |
| SAM, circular FPM | $5.1 \times 10^{-4}$ | $8.9 \times 10^{-3}$ | $8.0 \times 10^{-4}$ | 0.021 |
| SAM, bowtie FPM | $4.6 \times 10^{-4}$ | $8.0 \times 10^{-3}$ | $6.8 \times 10^{-4}$ | 0.017 |
| CLOWFS | $1.3 \times 10^{-3}$ | 0.022 | $2.3 \times 10^{-3}$ | 0.066 |

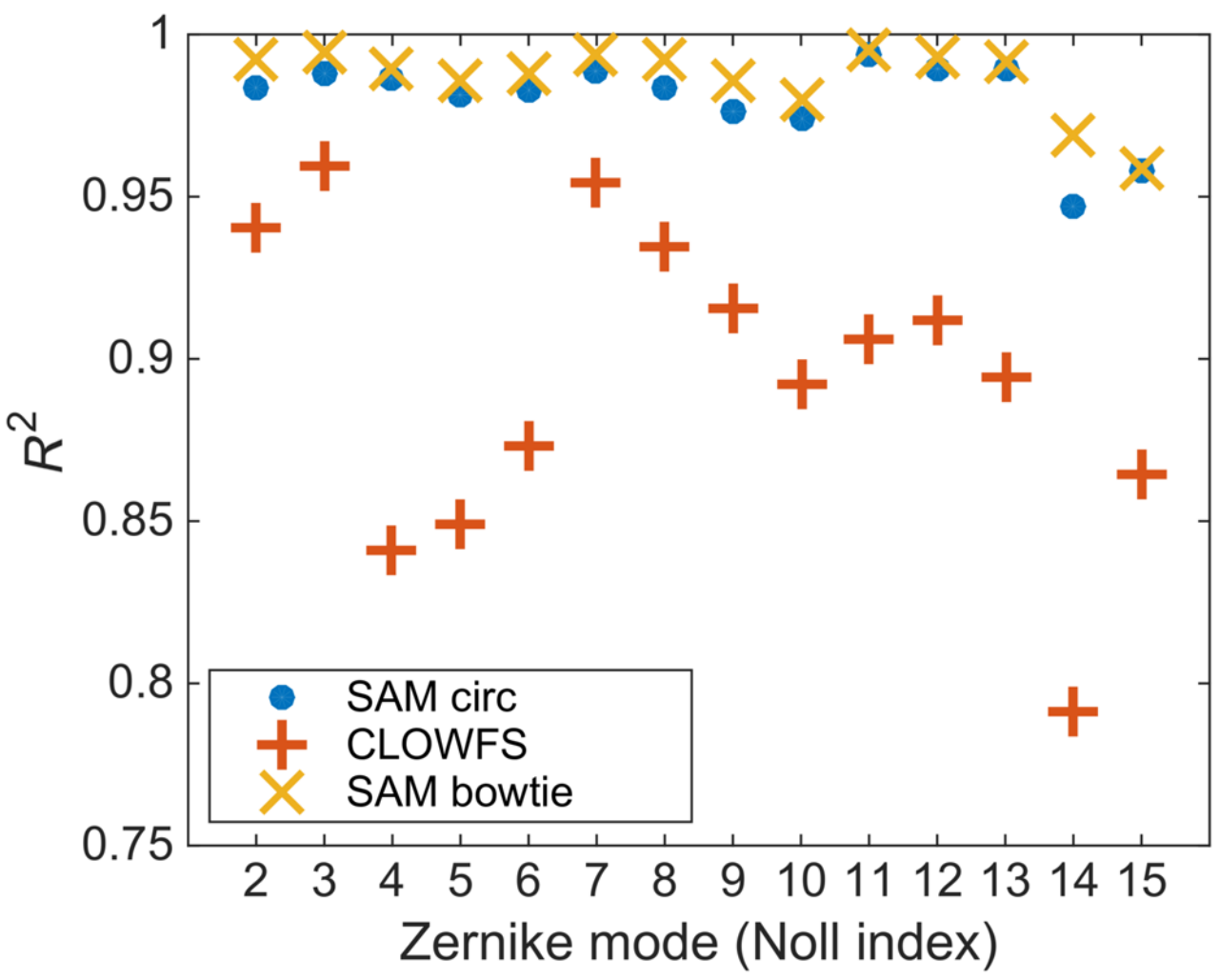

**Fig. 6** The $R^2$ fit metric of the SAM WFS and CLOWFS, averaged for each mode over 100 random wavefront realizations. The SAM results are plotted with both tested FPM shapes: circular and bowtie.

### 3.2 *Phase Reconstruction Tests*

We illustrate the ability to translate the coefficient estimates to a reconstructed phase distribution in Fig. 7. The input phase map in Fig. 7(a) comprised the same combination of $\lambda/30$ defocus and coma modes used to simulate the aberrated PSF and contrast curve in Fig. 1. The residual error of the SAM WFS estimate is $6 \times 10^{-3}$ waves peak to valley. We propagated this same residual phase map of Fig. 7(c) through the shaped pupil to show the recovery of the coronagraph contrast, resulting in the curve labeled "SAM WFS corrected" in Fig. 1(c).

We also considered the response of the system to three more cases: (1) wavefronts consisting of pure Zernike modes of varying amplitude, (2) a random phase screen obtained by a linear combinations of Zernike modes ($n = 1 - 4$), and (3) a Kolmogorov phase map.

For case (1), an increasing number of each Zernike mode is applied to a flat wavefront to analyze the response of the sensor to low-order aberrations [Fig. 8(a)]. The response of both WFSs is linear for Zernike coefficients up to $\lambda/20$ waves RMS, and deviates about 10% from linear near $\lambda/10$ waves RMS. When compared with a similar calculation published for a Zernike WFS,[51] our SAM and CLOWFS models exhibit a somewhat larger range of linearity upto $0.05\lambda$ waves RMS as compared to $0.03\lambda$ waves RMS for Zernike WFS.

For case (2), we created a phase map composed of a linear combination of Zernike modes with random coefficients, with the same constraints as the evaluations in Sec. 3.1. In Fig. 8(b), we plot the value of the 14 Zernike coefficients, comparing the true input values and the estimates side by side. At each mode, the coefficient estimate is accurate to within $\lambda/100$, consistent with Table 3.

The case of a Kolmogorov phase map is an interesting test for the concept, because in general we expect real optical surfaces to create a full spectrum of phase errors, with spatial frequency far above what the SAM WFS can estimate. Using a program to generate a random phase distribution with a Kolmogorov power spectrum (power law exponent −11/3), we tested several realizations with the resulting phase scaled up to $0.05\lambda$ RMS over the open area of the ripple SPC. One example of this reconstruction is shown in Fig. 9. The left-hand panel [Fig. 9(a)] shows the original phase map at the entrance of the SPC. The reconstructed phase [Fig. 9(b)] has $0.038\lambda$ RMS and there is $0.030\lambda$ RMS of residual high-order aberrations [Fig. 9(c)] after the estimated phase was subtracted from the input.

### 3.3 *Response to a Rapid Pointing Error*

We expect tip-tilt errors arising from line-of-sight pointing oscillations to be the most quickly varying aberration mode for a space telescope, and also one of the most important types of wavefront error to correct. Therefore, we assess the ability of SAM WFS and CLOWFS to sense a rapid pointing error in a regime of a fixed, realistic exposure time.

We assume the space telescope has an open, two-meter diameter circular aperture, and that the coronagraph operates at a central wavelength of $0.55\,\mu$m. The target star has a $V$-band apparent magnitude of 4.83, appropriate for a Sun-like star at a distance 10 pc. We collect light over a 20% bandwidth, and model this bandwidth in our Fourier propagation by averaging the sensor intensity pattern computed at five wavelengths spanning the passband. We assume losses due to reflections upstream of the coronagraph accumulate to 50% of the energy incident on the telescope primary, and a detector quantum efficiency of 0.8 $e^-$/photon.

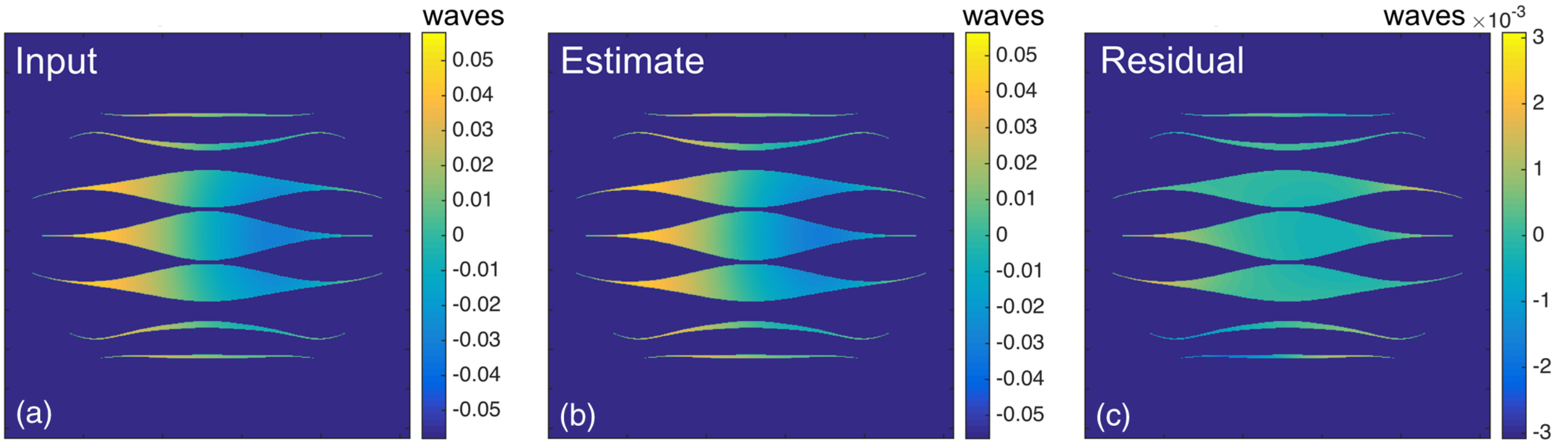

**Fig. 7** (a) Input phase map at pupil plane with a defocus and coma modes, each of amplitude $\lambda/30$ waves RMS; (b) the corresponding reconstructed phase from the SAM WFS model, and (c) residual error of the estimated phase, with peak-to-valley $6 \times 10^{-3}$ waves. In all plots, the phase is displayed in units of wavelength.

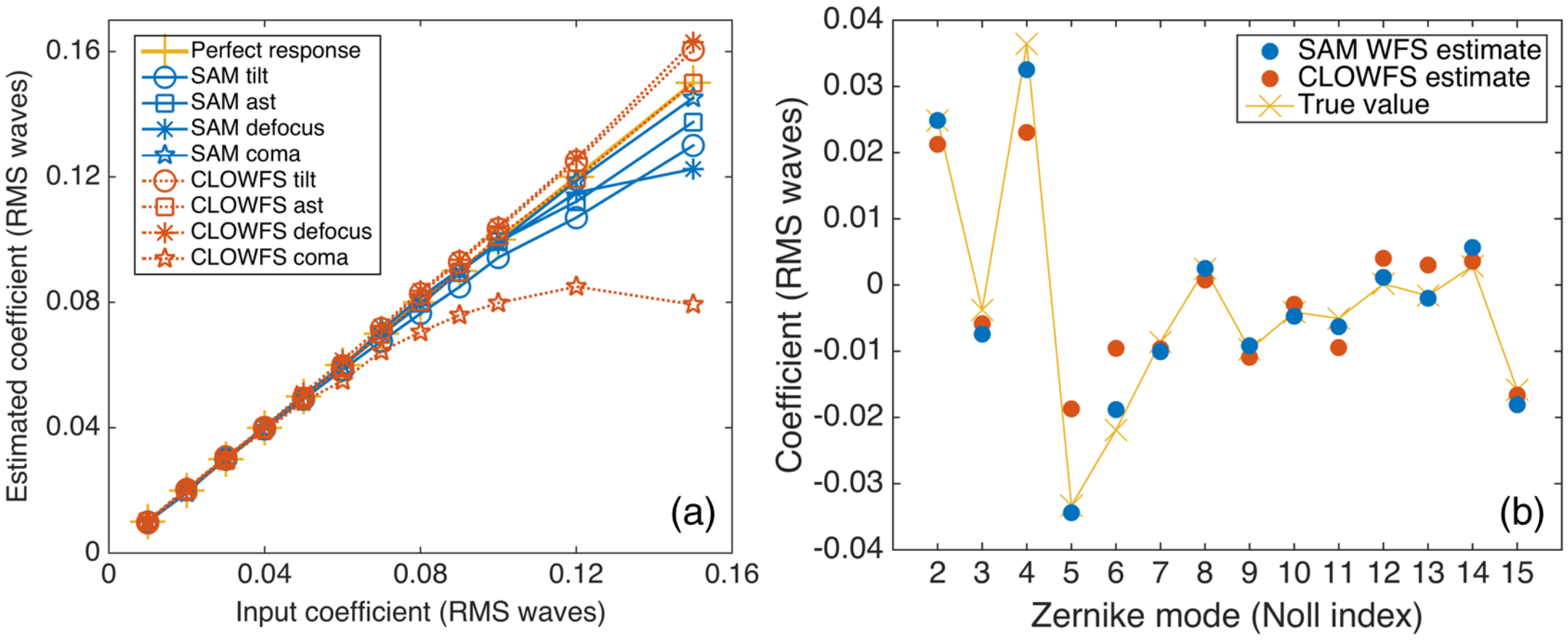

**Fig. 8** (a) Response of the SAM wavefront sensor to pure Zernike modes, showing the range of linearity, and (b) an example of the Zernike coefficient estimation for one random wavefront realization.

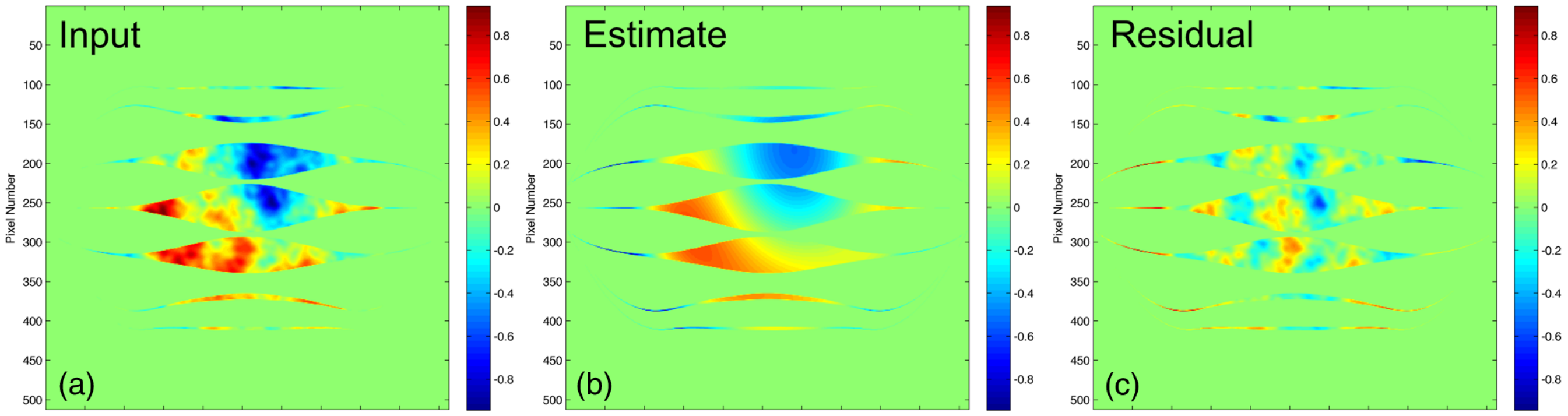

**Fig. 9** (a) Input phase map with a Kolmogorov power spectrum, (b) phase estimate for modes up to radial degree $n = 4$, and (c) the residual error after subtracting the estimate from the input. The phase is plotted here in units of radians.

The ripple shaped pupil transmits 14.7% of the incident energy relative to an open circular aperture, and the FPM mask (in this test fixed as a disk of radius $4\lambda/D$) transmits 56.7% of the energy incident on the coronagraph focal plane. For the SAM, due to the aperture mask there is an additional ratio of 9.1% between the energy arriving at the sensor and the energy reflected by the FPM. The distribution of the nominal intensity patterns of SAM and CLOWFS also differs significantly. The CLOWFS intensity is more centrally concentrated: for a CCD sampling of four pixels per $\lambda/D$, the ratio of energy in the peak pixel to the full image is 1.10%, versus 0.22% for the SAM WFS.

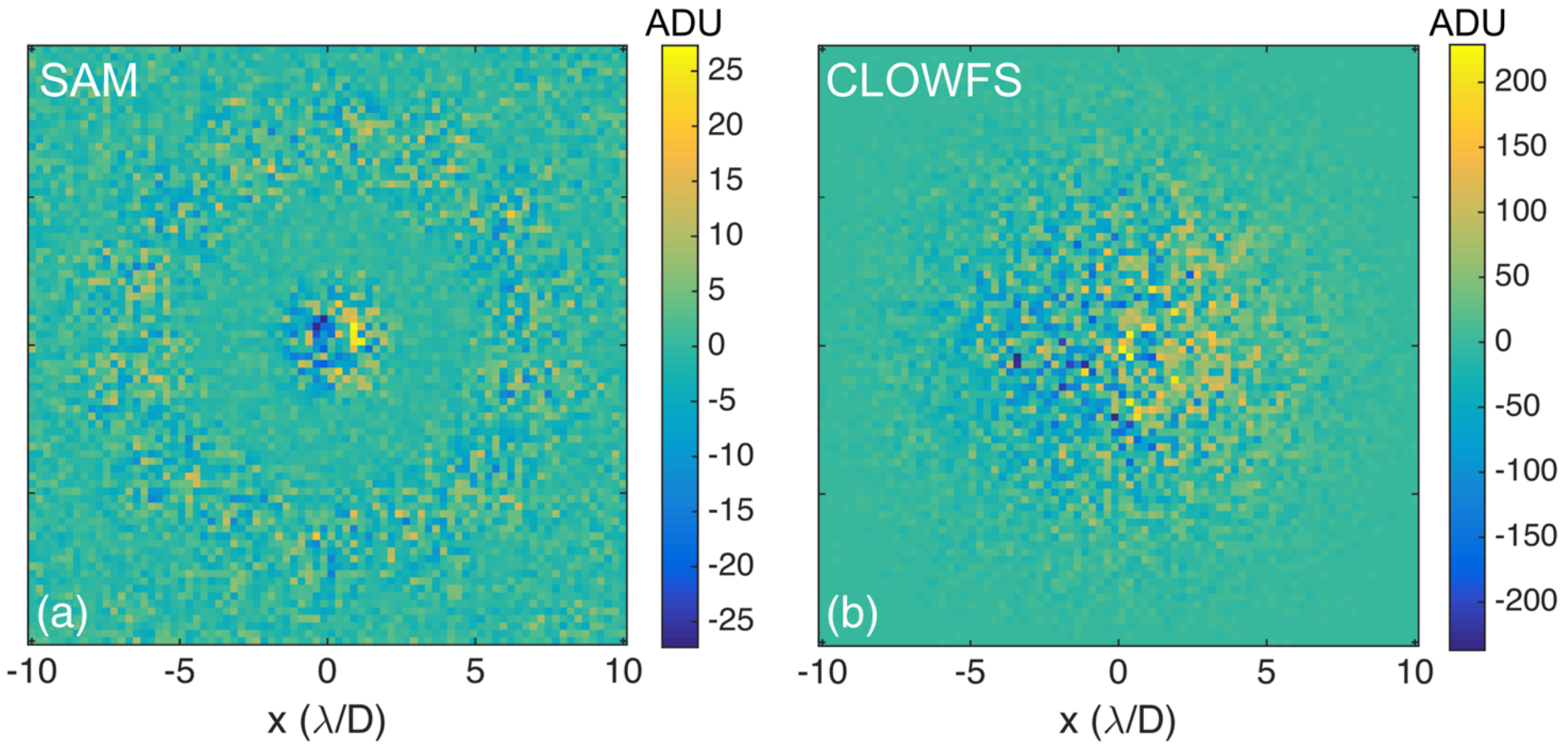

**Fig. 10** Simulated charge-coupled device (CCD) image differences between nominal and aberrated intensities for the (a) SAM and (b) CLOWFS wavefront sensors due to a 1 mas pointing error, observed with a 0.05 s integration time over a 20% bandwidth.

**Table 4** Standard deviations of the tilt estimates for 100 realizations of a 0.05 s exposure, for wavefronts with 1 and 10 mas pointing errors.

| | | Estimate standard deviation (milliarcsec) |
|---|---|---|
| 1 mas tilt | SAM WFS | 0.36 |
| | CLOWFS | 0.080 |
| 10 mas tilts | SAM WFS | 0.37 |
| | CLOWFS | 0.082 |

We simulate the differential wavefront tilt signal collected over a 0.05 s integration time for 1 and 10 milliarcsec (mas) pointing errors. The simulation model includes the diffraction propagation of five different wavelengths for 20% bandwidth centered at wavelength of 0.55 $\mu$m. This integration time is appropriate for controlling a jitter oscillation with a temporal frequency near or below 1 Hz. The phase gradient for a 1 mas pointing error is 0.0176 waves peak-to-valley across the telescope diameter. We simulate each CCD array with a read noise of standard deviation 1 ADU, although in practice this could be lower if the instrument used an electron-multiplying charge-coupled device (EMCCD) customized for low flux operation.[52] In Fig. 10, we show the noisy differential CCD image (aberrated minus nominal intensity) for the 1 mas pointing error, at resolution four pixels per $\lambda/D$. We tested the response of both WFSs to 100 signal realizations for each pointing error.

The lower throughput of the SAM WFS is partially compensated for by the fact that the differential tilt signal is proportionally larger to the peak intensity. For a 1 mas tilt, the peak-to-valley difference signal is 2.5% of the peak SAM intensity, versus 1.1% for the CLOWFS. However, after taking into account the throughput to the sensor and the distribution of the intensity, the difference signal for the wavefront tilt is still 13 times higher for CLOWFS than for SAM. For short exposure times with low signal-to-noise ratio, this naturally results in poorer estimation. For both sensors, the estimate distribution is centered on the true value. Therefore, in Table 4, we simply list the sample standard deviations to indicate the typical errors. The scatter in the SAM estimate is a factor of ∼5 wider for both tilt levels.

## 4 Discussion

The SAM WFS is best suited for estimating dynamic aberrations. The estimation algorithm relies on data acquired from a zero point meeting the target wavefront goal (which is not necessarily a flat phase, although here we have written our sensing equation in a way that assumes this). The initial reference point corresponding to the flat phase could be reached using phase diversity,[53] or another wavefront calibration method based in the science focal plane. Then, the SAM WFS response matrix could be measured using the DM. The differential approach to estimation reduces the inherent vulnerability to noncommon path errors. For a system with these limitations, a space coronagraph is the most compelling application, and was the primary motivation for this study. In this situation, the high-order spatial frequency corrections require long integration times due to the faint signal in the relevant region of the science focal plane. In order for the high-order wavefront control system of a space-based exoplanet imaging instrument to work effectively, the low-order aberrations need to be stabilized between deformable mirror correction steps. Therefore, the SAM WFS, like the CLOWFS and the phase contrast/Zernike WFS, could be used to estimate changes in the wavefront on a time scale much shorter than the high-order wavefront control loop. The SAM WFS also has potential uses for ground-based, extreme AO exoplanet imaging instruments, however. In the case where flexure-induced, noncommon path aberrations are a concern for the coronagraph performance, the SAM WFS could help to lock in on some initial low-order corrections by offsetting the DM commands of the AO system.

Throughout our analysis we used the CLOWFS as a reference point for performance, for two reasons: (1) that it is a relatively mature low-order WFS concept with strong experimental verification, and (2) because of the similarities in the optical configuration, a direct comparison required only straightforward modifications to the SAM WFS Fourier propagation model. A carefully planned comparison between models of a SAM WFS and a Zernike WFS would also be of great value, however. Future work will include comparison of all these three techinques. A study of their relative merits might take into account the effects of throughput, the properties of the differential signal, linearity, and chromaticity, for example.

In our comparisons with CLOWFS, we found that the accuracy of the SAM estimation is promising, but that the light loss through the aperture hinders the performance in estimation scenarios with short exposure time. We know, however, from trials with different aperture mask shapes that there is a tradeoff that can be explored between mask throughput and estimation accuracy. This is expected, since the diversity in the differential intensity signal, and therefore the ability to discriminate phase modes, to some extent depends on nonredundancy among the baselines joining the holes. This consideration tends to favor smaller samples at the re-imaged pupil.

Along these lines, we found that the condition number of the linear system matrix in Eq. (10) was a useful diagnostic of the estimation accuracy of a given mask shape. For example, the maximally redundant aperture, an open circle, has an extremely high condition number, because the corresponding system matrix is ill-conditioned. When we add an ad hoc asymmetric obscuration, such as the radial bar used by Martinache in his WFS,[46] the condition number becomes acceptable—below 1000, yet an order of magnitude above that of the sparse aperture mask presented here. If we shrink the holes of that sparse aperture mask, the condition number decreases. In the future, we will revisit the problem of optimizing the SAM geometry for low-order wavefront sensing, by applying numerical tools to find solutions that reach a compromise between accuracy and estimation speed.

## 5 Conclusion

Exquisite control of low-order wavefront aberrations is necessary for a coronagraph to reach its designed performance. We presented a coronagraph-integrated WFS, relying on a SAM to infer dynamic low-order aberrations from a differential intensity distribution of re-imaged, discarded starlight. We proposed a simple algorithm to sense the coefficients of the low-order aberrations and verified it with a Fourier propagation model. Wavefront estimation accuracy has been quantified for pure Zernike modes, random linear combinations of Zernike modes, and a phase screen with a Kolmogorov power spectrum. We find that the SAM WFS can sense and discriminate all low-order aberrations (Zernike modes with radial degree $n = 1 - 5$).

Although in this paper we only predicted the performance for a particular shaped pupil coronagraph, the SAM WFS can be adapted to any coronagraph with an opaque focal plane mask, such as an APLC. Future work will be directed at verifying our simulations with laboratory experiments and numerically optimizing the shape of the SAM for a given coronagraph and wavefront estimation problem.

## Acknowledgments

We thank James B. Breckinridge for his useful suggestions on the geometry of the sparse aperture mask (SAM). We benefitted from several discussions with Tyler Groff during the development of the concept. We also thank J. Kent Wallace and Fang Shi of the Jet Propulsion Laboratory for their early feedback on the manuscript. This work was partially funded by NASA Technology Development for Exoplanet Missions (TDEM) under Grant No. NNX09AB96G and by the Jet Propulsion Laboratory of the California Institute of Technology. A. J. E. Riggs is supported by NASA under Grant No. NNX14AM06H, a NASA Space Technology Research Fellowship.

## References

1. A. J. Skemer et al., "First light LBT AO Images of HR 8799 bcde at 1.6 and 3.3 $\mu$m: new discrepancies between young planets and old brown dwarfs," *Astrophys. J.* **753**, 14 (2012).
2. B. R. Oppenheimer et al., "Reconnaissance of the HR 8799 exosolar system. I. Near-infrared spectroscopy," *Astrophys. J.* **768**, 24 (2013).
3. Q. M. Konopacky et al., "Detection of carbon monoxide and water absorption lines in an exoplanet atmosphere," *Science* **339**, 1398–1401 (2013).
4. M. Janson et al., "Direct imaging detection of methane in the atmosphere of GJ 504 b," *Astrophys. J.* **778**, L4 (2013).
5. N. J. Kasdin et al., "Extrasolar planet finding via optimal apodized-pupil and shaped-pupil coronagraphs," *Astrophys. J.* **582**, 1147–1161 (2003).
6. http://wfirst.gsfc.nasa.gov://wfirst.gsfc.nasa.gov
7. A. Carlotti, N. J. Kasdin, and R. J. Vanderbei, "Shaped pupil coronagraphy with WFIRST-AFTA," *Proc. SPIE* **8864**, 886410 (2013).
8. A. J. E. Riggs et al., "Shaped pupil design for future space telescopes," *Proc. SPIE* **9143**, 914325 (2014).
9. I. Poberezhskiy et al., "Technology development towards WFIRST-AFTA coronagraph," *Proc. SPIE* **9143**, 91430P (2014).
10. A. Sivaramakrishnan et al., "Sensing phase aberrations behind Lyot coronagraphs," *Astrophys. J.* **688**, 701–708 (2008).
11. O. Guyon, T. Matsuo, and R. Angel, "Coronagraphic low-order wavefront sensor: principle and application to a phase-induced amplitude coronagraph," *Astrophys. J.* **693**, 75–84 (2009).
12. F. Malbet, J. W. Yu, and M. Shao, "High-dynamic-range imaging using a deformable mirror for space coronagraphy," *Publ. Astron. Soc. Pac.* **107**, 386 (1995).
13. P. J. Bordé and W. A. Traub, "High-contrast imaging from space: speckle nulling in a low-aberration regime," *Astrophys. J.* **638**, 488–498 (2006).
14. F. Martinache et al., "On-sky speckle nulling demonstration at small angular separation with SCExAO," *Publ. Astron. Soc. Pac.* **126**, 565–572 (2014).
15. A. Give'on et al., "Broadband wavefront correction algorithm for high-contrast imaging systems," *Proc. SPIE* **6691**, 66910A (2007).
16. L. Pueyo et al., "optimal dark hole generation via two deformable mirrors with stroke minimization," *Appl. Opt.* **48**, 6296–6312 (2009).
17. T. D. Groff and N. Jeremy Kasdin, "Kalman filtering techniques for focal plane electric field estimation," *J. Opt. Soc. Am. A* **30**, 128 (2013).
18. J.-F. Sauvage et al., "Calibration and precompensation of noncommon path aberrations for extreme adaptive optics," *J. Opt. Soc. Am. A* **24**, 2334–2346 (2007).
19. B. Paul, J.-F. Sauvage, and L. M. Mugnier, "Coronagraphic phase diversity: performance study and laboratory demonstration," *Astron. Astrophys.* **552**, A48 (2013).
20. T. D. Groff et al., "Wavefront control scenarios for a coronagraph on an AFTA-like space telescope," *Proc. SPIE* **8864**, 886413 (2013).
21. B. R. Oppenheimer et al., "The Lyot project: toward exoplanet imaging and spectroscopy," *Proc. SPIE* **5490**, 433–442 (2004).
22. A. P. Digby et al., "The challenges of coronagraphic astrometry," *Astrophys. J.* **650**, 484–496 (2006).
23. F. P. A. Vogt et al., "Coronagraphic low-order wavefront sensor: post-processing sensitivity enhancer for high-performance coronagraphs," *Publ. Astron. Soc. Pac.* **123**, 1434–1441 (2011).
24. K. Cavanagh, "Analysis and development of a low-order wavefront sensor for exoplanet detection applications," Senior Thesis, Princeton University (2014).
25. J. K. Wallace et al., "Science camera calibration for extreme adaptive optics," *Proc. SPIE* **5490**, 370–378 (2004).
26. J. L. Codona and R. Angel, "Imaging extrasolar planets by stellar halo suppression in separately corrected color bands," *Astrophys. J.* **604**, L117–L120 (2004).
27. P. Baudoz et al., "Direct imaging of exoplanets: science and techniques," *IAU Colloq.* **200**(1), 553–558 (2005).
28. R. Galicher, P. Baudoz, and G. Rousset, "Wavefront error correction and Earth-like planet detection by a self-coherent camera in space," *Astron. Astrophys.* **488**, L9–L12 (2008).
29. F. Zernike, "Phase contrast, a new method for the microscopic observation of transparent objects," *Physica* **9**(7), 686–698 (1942).
30. R. H. Dicke, "Phase-contrast detection of telescope seeing errors and their correction," *Astrophys. J.* **198**, 605–615 (1975).
31. E. E. Bloemhof and J. K. Wallace, "Phase contrast techniques for wavefront sensing and calibration in adaptive optics," *Proc. SPIE* **5169**, 309–320 (2003).
32. J. K. Wallace et al., "Phase-shifting Zernike interferometer wavefront sensor," *Proc. SPIE* **8126**, 81260F (2011).
33. M. N'Diaye et al., "Design optimization and lab demonstration of ZELDA: a Zernike sensor for near-coronagraph quasi-static measurements," *Proc. SPIE* **9148**, 91485H (2014).
34. J. W. Goodman, *Introduction to Fourier Optics*, 3rd ed., Roberts and Company Publishers, Greenwood Village, CO (2005).
35. P. G. Tuthill, J. D. Monnier, and W. C. Danchi, "A dusty pinwheel nebula around the massive star WR104," *Nature* **398**, 487–489 (1999).
36. M. J. Ireland et al., "Dynamical mass of GJ 802B: a brown dwarf in a triple system," *Astrophys. J.* **678**, 463–471 (2008).
37. S. Hinkley et al., "Observational constraints on companions inside of 10 AU in the HR 8799 planetary system," *Astrophys. J.* **730**, L21 (2011).
38. J. E. Baldwin et al., "Closure phase in high-resolution optical imaging," *Nature* **320**, 595–597 (1986).
39. C. A. Haniff et al., "The first images from optical aperture synthesis," *Nature* **328**, 694–696 (1987).
40. P. G. Tuthill et al., "Michelson interferometry with the Keck I Telescope," *Publ. Astron. Soc. Pac.* **112**, 555–565 (2000).
41. M. J. Ireland, "Phase errors in diffraction-limited imaging: contrast limits for sparse aperture masking," *Mon. Not. R. Astron. Soc.* **433**, 1718–1728 (2013).
42. P. Tuthill et al., "Sparse-aperture adaptive optics," *Proc. SPIE* **6272**, 62723A (2006).
43. A. Z. Greenbaum et al., "Gemini planet imager observational calibrations X: non-redundant masking on GPI," *Proc. SPIE* **9147**, 91477B (2014).
44. A. Sivaramakrishnan et al., "Non-redundant masking ideas on JWST," *Proc. SPIE* **9143**, 91433S (2014).
45. A. Sivaramakrishnan et al., "Non-redundant aperture masking interferometry (AMI) and segment phasing with JWST-NIRISS," *Proc. SPIE* **8442**, 84422S (2012).
46. F. Martinache, "The asymmetric pupil Fourier wavefront sensor," *Publ. Astron. Soc. Pac.* **125**, 422–430 (2013).
47. F. Martinache et al., "Wavefront control scenarios for a coronagraph on an AFTA-like space telescope," *Proc. SPIE* **9148**, 914821 (2014).
48. B. Pope et al., "Wavefront sensing from the image domain with the Oxford-SWIFT integral field spectrograph," *Proc. SPIE* **9148**, 914859 (2014).
49. N. J. Kasdin, R. J. Vanderbei, and R. Belikov, "Shaped pupil coronagraphy," *C. R. Phys.* **8**, 312–322 (2007).

50. N. T. Zimmerman, "High-contrast observations with an integral field spectrograph," PhD Thesis, Columbia University (2011).
51. M. N'Diaye et al., "Calibration of quasi-static aberrations in exoplanet direct-imaging instruments with a Zernike phase-mask sensor," *Astron. Astrophys.* **555**, A94 (2013).
52. N. Smith et al., "EMCCD technology and its impact on rapid low-light photometry," *Proc. SPIE* **5499**, 162–172 (2004).
53. R. G. Paxman, T. J. Schulz, and J. R. Fienup, "Joint estimation of object and aberrations by using phase diversity," *J. Opt. Soc. Am. A* **9**, 1072 (1992).

**Hari Subedi** received his BS degree in aerospace engineering and mathematics from the University of Arizona in 2013. Currently, he is in his second year of graduate study at the Department of Mechanical and Aerospace Engineering at Princeton University.

**Neil T. Zimmerman** is a postdoctoral research associate in the Department of Mechanical and Aerospace Engineering at Princeton University. After studying electrical engineering at Cooper Union, he received his PhD degree in astronomy from Columbia University in 2011. He is a member of SPIE and the American Astronomical Society, and specializes in instrumentation, observational methods, and data processing techniques for exoplanet imaging.

**N. Jeremy Kasdin** is a professor of mechanical and aerospace engineering and Vice Dean of the School of Engineering and Applied Science at Princeton University. He is the principal investigator of Princeton's High Contrast Imaging Laboratory. He received his PhD degree from Stanford University in 1991. His research interests include space systems design, space optics and exoplanet imaging, orbital mechanics, guidance and control of space vehicles, optimal estimation, and stochastic process modeling. He is an associate fellow of the American Institute of Aeronautics and Astronautics, and member of the American Astronomical Society and SPIE.

**Kathleen Cavanagh** received her BSE degree in mechanical and aerospace engineering from Princeton University in 2014. Since graduating, she has been working as part of a small engineering team at SciTec, Inc., to research and develop advanced and rapid multisensor multitarget state estimation and tracking capabilities.

**A J Eldorado Riggs** is a PhD candidate in the Department of Mechanical and Aerospace Engineering at Princeton University. He received his BS degree in physics and mechanical engineering from Yale University in 2011. His research interests include wavefront estimation and control and coronagraph design for exoplanet imaging.